\documentclass[twocolumn,aps,showpacs,preprintnumbers,superscriptaddress]{revtex4}
\usepackage{graphicx}
\usepackage{dcolumn}
\usepackage{bm}

\usepackage{amsmath}

\begin{document}

\title{Classical-interference analog of quantum fluctuations for bound-state
soliton pairs}
\author{Ray-Kuang Lee}
\affiliation{Department of Photonics and Institute of Electro-Optical Engineering,
National Chiao-Tung University, Hsinchu 300, Taiwan}
\author{Yinchieh Lai}
\affiliation{Department of Photonics and Institute of Electro-Optical Engineering,
National Chiao-Tung University, Hsinchu 300, Taiwan}
\email{yclai@mail.nctu.edu.tw}
\author{Boris A. Malomed}
\affiliation{Department of Interdisciplinary Studies, Faculty of Engineering, Tel Aviv
University, Tel Aviv 69978, Israel}

\begin{abstract}
Quantum photon-number fluctuation and correlation of bound soliton pairs in mode-locked fiber lasers are studied based on the complex Ginzburg-Landau equation model. 
We find that, depending on their phase difference, the total photon-number noise of the bound soliton pair can be larger or smaller than that of a single soliton and the two solitons in the soliton pairs are with positive or negative photon-number correlation, correspondingly. It is predicted for the first time that out-of-phase soliton pairs can exhibit less noises due to negative correlation.
\end{abstract}
\maketitle
\date{\today}

%\ocis{270.5530, 140.3510, 140.4050.}

Quantum solitons have attracted a great deal of research interest in the contexts of nonlinear quantum optics, condensed-matter physics, and quantum information science due to their remarkable nonclassical properties.
In particular, quantum solitons in optical fibers largely resemble their classical counterparts, but with additional quantum fluctuations around the mean fields.
It has been possible to achieve \textit{squeezing} through quantum solitons in optical fibers, \cite{Carter87, Drummond87, Lai89a, Lai89b} and they may also serve as a new platform for quantum information applications \cite{Silberhorn01, Silberhorn02, Konig02}.

Quantum solitons are macroscopic optical wave packets which offer a testbed for quantum optics and quantum field theories. For the quantum nonlinear Schr{\"{o}}dinger equation (NLSE), exact soliton states can be constructed as combinations of eigenstates of the Hamiltonian of the one-dimensional Bose gas with $\delta $-like (contact) interaction through the Bethe ansatz method \cite{Lai89b}.
In the large photon number limit, which corresponds to the usual optical solitons generated by lasers, the many photon wave function of the quantum soliton is well approximated by a single-photon wave function (the Hartree approximation) \cite{Lai89a}. Linearization around such a soliton \cite{Haus90, Lai93}
successfully explains experimental observations of quantum fluctuations for temporal fiber solitons, provided that optical loss and higher-order effects are negligible \cite{Rosenbluh, Bergman91, Friberg, Krylov99, Spalter}.

It is well known that the force between adjacent solitons in the NLSE model is attractive or repulsive, depending on the phase difference between them \cite{Agrawal95}. Stationary bound soliton states in this conservative model do not exist. Formation of effectively stable double-, triple-, and multi-soliton bound states was predicted in models based on the complex Ginzburg-Landau equation (CGLE) \cite{Malomed91, Akhmediev96, Soto-Crespo03}, and observed experimentally in various passively mode-locked fiber lasers \cite{Tang01, Seong02,Grelu03}. The separation between the solitons in these bound states are ``quantized", taking a set of discrete values. The amplitude noise in triplet bound states generated by a stretched-pulse ytterbium-doped double-clad fiber laser was observed to be reduced compared to the single soliton pulse \cite{Ortac04}. It is an issue of straightforward interest to study the noise of these bound solitons, and to understand why the mode-locked fiber lasers operate more stably in the bound-state regime.
\begin{figure}[b]
\centerline{\includegraphics[width=3.0in, height=2.2in]{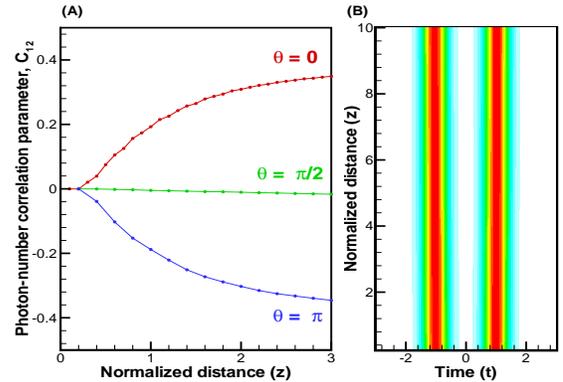}}
\caption{(A) The photon-number correlation parameter, $C_{12}$,
for the bound-state soliton pairs with different relative phases.
(B) The contour plot of the classical solution for the bound
state. All the results are presented for $D=1$, $\protect\delta
=-0.01$, $\protect\epsilon =1.8$, $\protect\beta =0.5$,
$\protect\mu =-0.05$, and $\protect\nu =0$ in the CGLE
(\protect\ref{eq_CGLE}).}\label{fig-c12}
\end{figure}

The passively mode-locked fiber lasers are quite accurately described by the
cubic-quintic CGLE. In a normalized form, the equation is
\begin{eqnarray}
iU_{z}+(D/2)U_{tt}+|U|^{2}U &=&i\delta U+i\epsilon |U|^{2}U+i\beta U_{tt}
\nonumber \\
&+&i\mu |U|^{4}U-\nu |U|^{4}U,  \label{eq_CGLE}
\end{eqnarray}where $U$ is the local amplitude of the electromagnetic wave, $z$ is the
propagation distance, $t$ is the retarded time, and $D=+1$ and $-1$
correspond, respectively, to the anomalous and normal dispersion. Besides
the group-velocity dispersion (GVD) and the Kerr effect, which are accounted
for by conservative terms on the left-hand side of Eq. (\ref{eq_CGLE}), the
model also includes the quintic correction to the Kerr nonlinearity, through
the coefficient $\nu $, and non-conservative terms. The coefficients $\delta
$, $\epsilon $, $\mu $, and $\beta $ account for the linear, cubic, and
quintic loss or gain, and spectral filtering, respectively.

In the CGLE model, with suitable parameters degenerate bound-state soliton pairs are known to exist
through the balance between the gain and loss, in the form
\cite{Malomed91,Akhmediev96}, $U(z,t)=U_{0}(z,t+\rho
)e^{-i\theta /2}+U_{0}(z,t-\rho )e^{i\theta /2}$, \ where $U_{0}$
is a single soliton solution, and $\rho $ and $\theta $ are the
separation and phase difference between the solitons. In this
Letter, we focus on the consideration of three fundamentally
different cases, corresponding to the bound states with the same
separation and amplitude, and $\theta =0$, $\pi /2$, and $\pi $
(the in-phase, orthogonal, and out-of-phase pair), respectively.

We compute the quantum fluctuations of these soliton pairs by dint of a
numerically implemented \textit{back-propagation method}
\cite{Lai95}, which may be summarized as follows. First of all,
we replace the classical function $U(z,t)$ in Eq.
(\ref{eq_CGLE}) by the quantum-field operator variable,
$\hat{U}(z,t)$, which satisfies the equal-coordinate Bosonic
commutation relations. Next, the equation is linearized around the
classical solution through the substitution of
$\hat{U}(z,t)=U_{0}(z,t)+\hat{u}(z,t)$, assuming large photon numbers
in the solitons. Then, a zero-mean additional noise operator,
$\hat{n}(z,t)$, is introduced to make the quantum perturbation
fields in the linearized equation satisfy the Bosonic
communication relations (see Ref. [\ref{ref-RK04-bd}] for more
details). By imposing suitable correlation functions for the noise operator, the minimum quantum noise in the considered nonconservative model is introduced. Therefore the results presented here represent a lower limit required by the fundamental principles of quantum mechanics.
\begin{figure}[t]
\centerline{\includegraphics[width=2.5in]{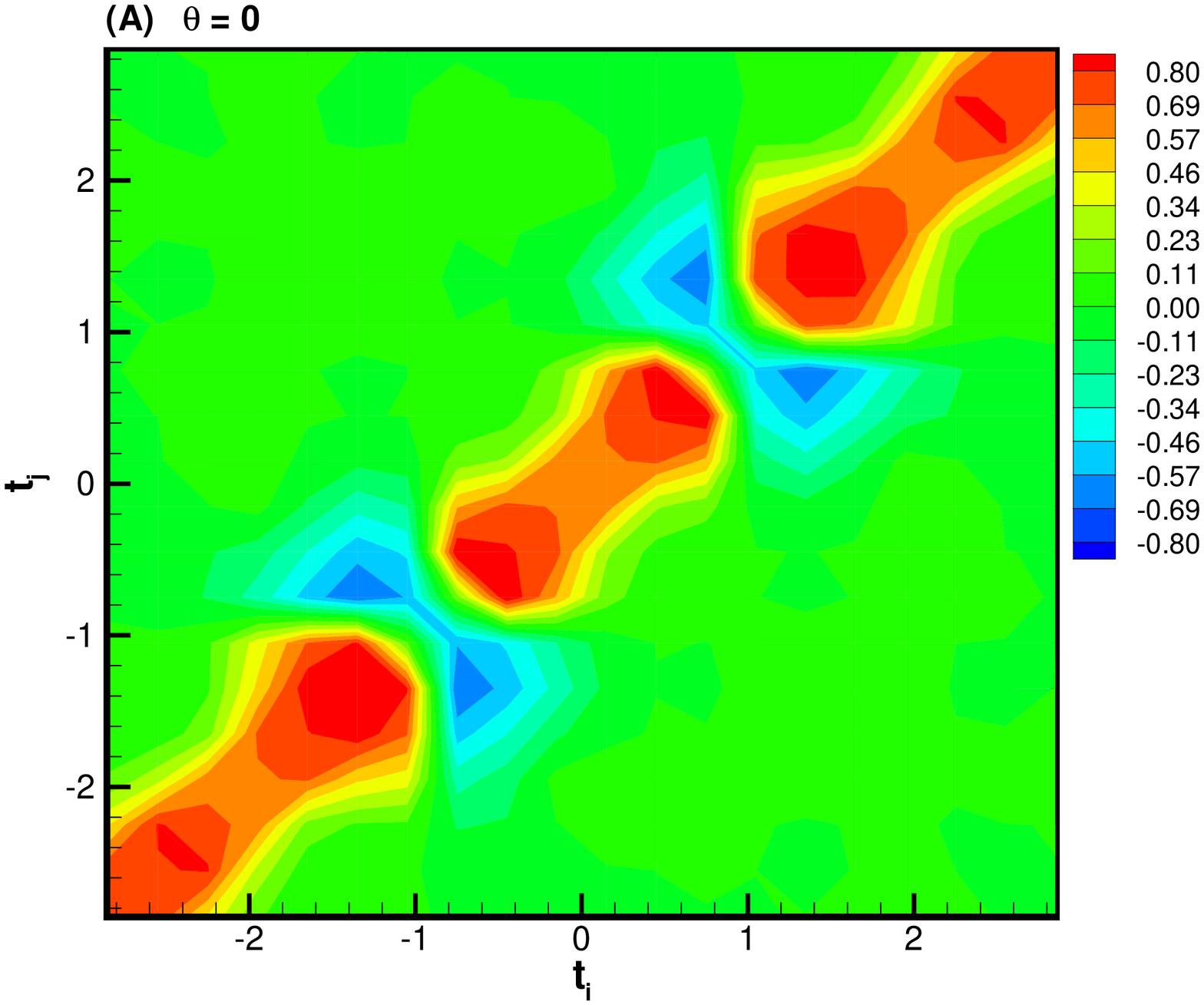}}
\centerline{\includegraphics[width=2.5in]{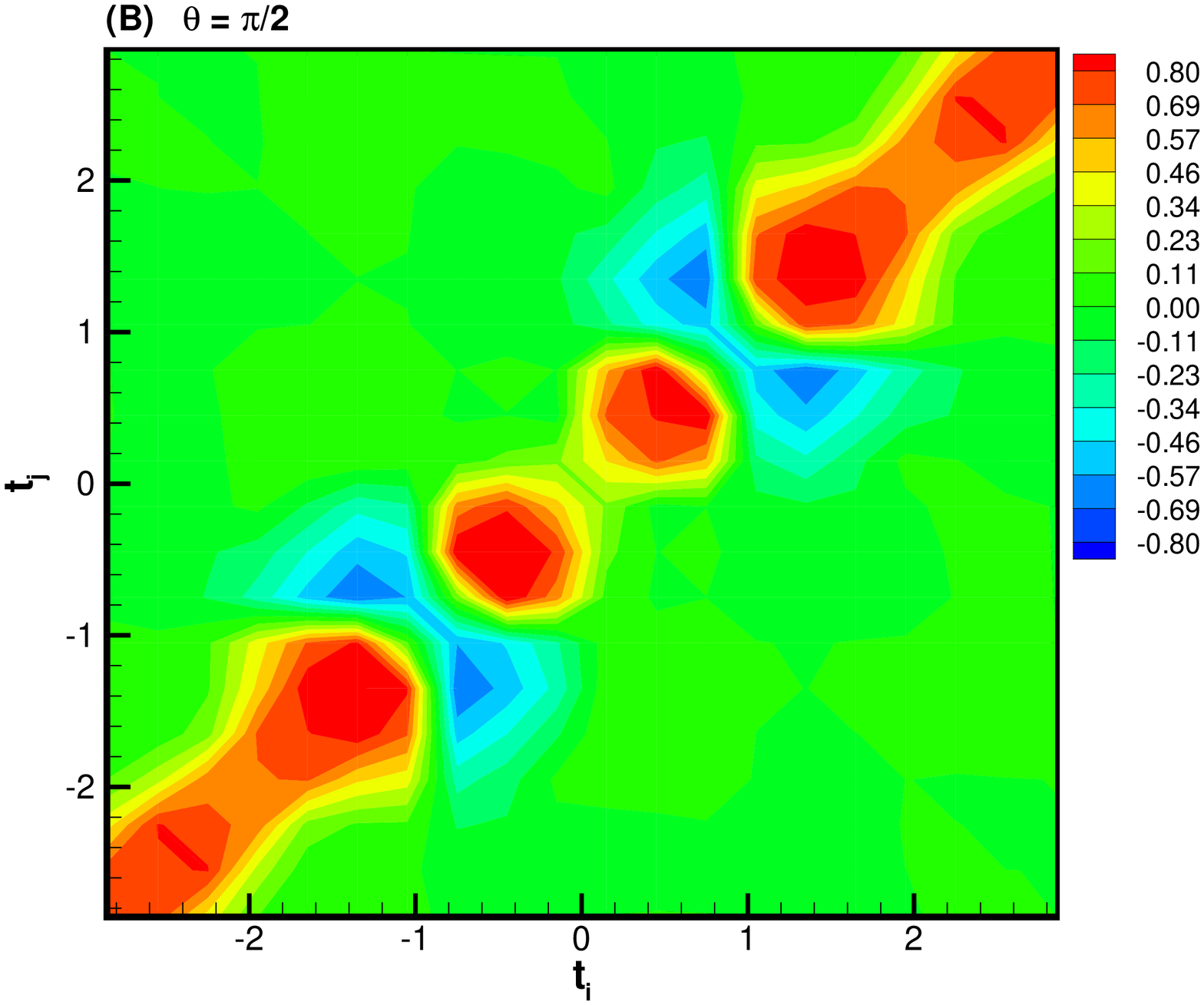}}
\centerline{\includegraphics[width=2.5in]{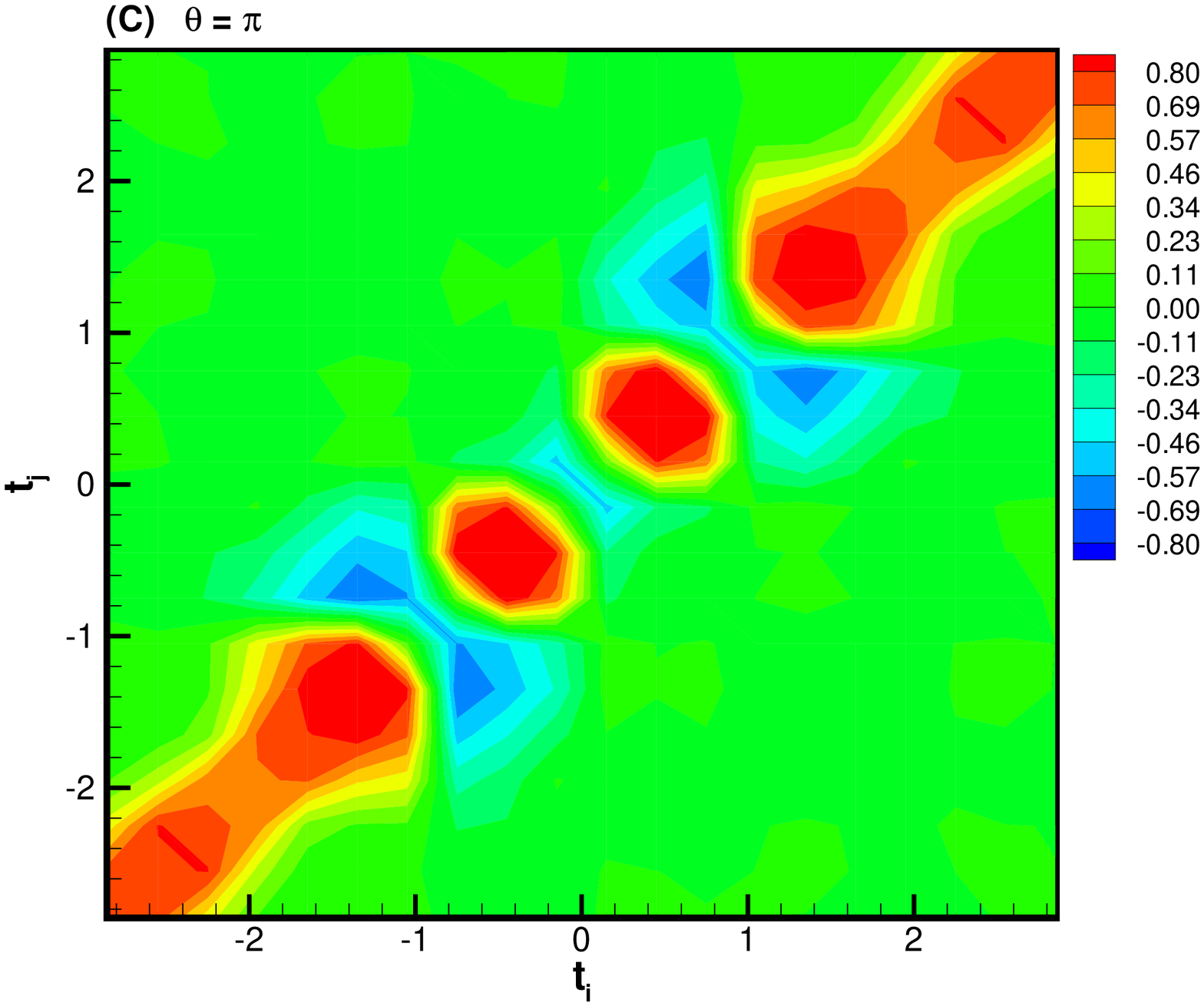}}
\caption{The time-domain photon-number correlation patterns,
$\protect\eta _{ij}$, for the bound soliton pairs with different
relative phases, after the normalized propagation distance
$z=0.4$. (A): $\protect\theta =0$, (B): $\protect\theta
=\protect\pi /2$, and (C): $\protect\theta =\protect\pi $. The time-division length $\Delta t = 0.3$}\label{fig-cij}
\end{figure}

Figure \ref{fig-c12} shows the photon-number \textit{correlation parameter}
for the two solitons in the bound soliton pair, which is defined as
\[
C_{12}=\frac{\langle :\Delta \hat{N}_{1}\Delta \hat{N}_{2}:\rangle
}{\sqrt{\langle \Delta \hat{N}_{1}^{2}\rangle \langle \Delta
\hat{N}_{2}^{2}\rangle }}.
\]Here, the colons stand for the normal ordering of the operators and  $\Delta
\hat{N}_{1,2}$ are perturbations of the photon-number operators for the two
solitons, which are numbered (1,2) according to their position in the time
domain. Initially, the two solitons are assumed to be uncorrelated,
with fluctuations around each soliton obeying the coherent-state statistics.
For the in-phase pair, the photon-number correlation
between the solitons gradually increases to positive values and eventually saturates around $C_{12}=0.36$.
But for the out-of-phase pair, $C_{12}$ gradually decreases to negative
values and then saturates too. In between, the correlation
parameter for the case of $\theta =\pi /2$ remains close to zero
as long as the computation is run. For the former two cases, the saturation of the photon-number
correlation parameter is due to the nonconservative effects in the CGLE
model.
\begin{figure}[t]
\centerline{\includegraphics[width=2.5in]{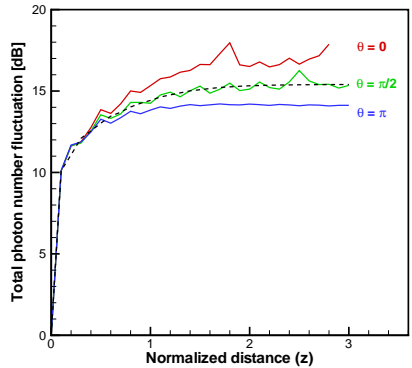}}
\caption{Comparison of the total photon-number fluctuations in the bound
soliton pairs with different relative phases (three solid lines), and around
the single soliton (the dash line).}
\label{fig-num}
\end{figure}

To further demonstrate the behavior difference of the photon-number correlation
for soliton pairs with different relative phases, in Fig. \ref{fig-cij} we display the \textit{time-domain} photon-number correlation patterns for them. The plotted correlation coefficients, $\eta _{ij}$,
are defined through the normally-ordered covariance,
\begin{equation}
\eta _{ij}\equiv \frac{\langle :\Delta \hat{n}_{i}\Delta \hat{n}_{j}:\rangle
}{\sqrt{\Delta \hat{n}_{i}^{2}\Delta \hat{n}_{j}^{2}}}~,  \label{C}
\end{equation}where $\Delta \hat{n}_{j}$ is the photon-number fluctuation in the $j$-th
time slot $\Delta t_{j}$,
\[
\Delta \hat{n}_{j}=\int_{\Delta t_{j}}d\,t[U_{0}(z,t)\hat{u}^{\dag
}(z,t)+U_{0}^{\ast }(z,t)\hat{u}(z,t)].
\]
Here the integral is taken over the given time slot, with the same time-division length $\Delta t$. Clearly, in Fig. \ref{fig-cij} (A) one can see that there
is a strong \emph{positive}-correlation band connecting the
quantum correlation patterns of the bound solitons when they are
in phase, $\theta =0$. In Fig. \ref{fig-cij} (C) there exists a
\emph{negative}-correlation pattern between two solitons for the
out-of-phase case, $\theta =\pi $. 
Moreover, for the case of $\theta =\pi /2$, in Fig. \ref{fig-cij} (B), the correlation
patterns of bound solitons are almost isolated.
In classical physics, in-phase and out-of-phase fields will lead
respectively to the constructive and destructive interference.
Here we observe a similar effect for the quantum noises.
What is more important, in Fig. \ref{fig-num} we compute the total photon number noise of the
bound soliton pair and compare it to the case of a single soliton (these results are
amenable to straightforward experimental verification). As one may expect,
the photon-number noise of the in-phase pair is larger than that for the
single soliton, which may be explained as the fluctuation enhancement due to
constructive interference. On the other hand, the noise is reduced for
the case of out-of-phase pair as the result of destructive
interference. The orthogonal soliton pair with $\theta =\pi /2$ may be viewed, in the first approximation, as independent two single solitons, which explains why it features almost the same noise level as the
single soliton, even though small oscillation of the noise level originated from the residual interaction between the two solitons can still be seen.

In conclusion, we have presented theoretical results on the
photon-number correlation and total photon-number noise for bound-state soliton pairs in
the model of complex cubic-quintic Ginzburg-Landau equation. The cases of the in-phase, orthogonal, and out-of-phase
soliton pairs have been considered in detail. We conclude that the
interference of the quantum fluctuations in the soliton pair is
constructive or destructive depending on the \emph{classical}
relative phase of the solitons. An important consequence of the results is
that the operation regime of the fiber laser should be more stable
when it is based on the \emph{out-of-phase} soliton pairs.

\end{document}